# Universal phase dynamics in VO₂ switches revealed by ultrafast *operando* diffraction

Aditya Sood[1,2,*], Xiaozhe Shen[3], Yin Shi[4], Suhas Kumar[5], Su Ji Park[3], Marc Zajac[2], Yifei Sun[6], Long-Qing Chen[4], Shriram Ramanathan[6], Xijie Wang[3], William C. Chueh[2], and Aaron M. Lindenberg[1,2,3,*]

[1]Stanford Institute for Materials and Energy Sciences, SLAC National Accelerator Laboratory, Menlo Park, CA 94025, USA.

[2]Department of Materials Science and Engineering, Stanford University, Stanford, CA 94305, USA.

[3]SLAC National Accelerator Laboratory, Menlo Park, CA 94025, USA.

[4]Department of Materials Science and Engineering, The Pennsylvania State University, University Park, PA 16802, USA.

[5]Hewlett Packard Labs, 1501 Page Mill Rd, Palo Alto, CA 94304, USA.

[6]School of Materials Engineering, Purdue University, West Lafayette, IN 47907, USA.

[*]Correspondence to aditsood@stanford.edu, aaronl@stanford.edu

**Strongly correlated materials that exhibit an insulator-metal transition are key candidates in the search for new computing platforms. Understanding the pathways and timescales underlying the electrically-driven insulator-metal transition is crucial for uncovering the fundamental limits of device operation. Using stroboscopic electron diffraction, we perform synchronized time-resolved measurements of atomic motions and electronic transport in operating vanadium dioxide switches. We discover an electrically-triggered, isostructural state that forms transiently on microsecond timescales, stabilized by local heterogeneities and interfacial interactions between the equilibrium phases. This metastable phase bears striking similarity to that formed under photoexcitation within picoseconds, suggesting a universal transformation pathway across eight orders of magnitude of timescale. Our results establish a new route for uncovering non-equilibrium and metastable phases in correlated materials, and open avenues for engineering novel dynamical behavior in nanoelectronics.**

The insulator-metal transition (IMT) in correlated oxide semiconductors is a striking example of an emergent phenomenon arising from the complex interplay between lattice and electronic degrees of freedom. Electronic and optical properties change drastically across the IMT, motivating applications in computing and photonics(*1*). In vanadium dioxide (VO₂), an archetypal correlated material, the IMT can be driven primarily in three ways: through thermal(*2–4*), optical(*5–11*), and electrical(*12–14*) excitation. Among these, the *electrically-triggered* IMT (E-IMT) is arguably the most sought after for future solid-state devices. It has been employed to create novel applications including steep sub-Boltzmann switching transistors(*15*), neuromorphic

circuits(*16, 17*), and reconfigurable photonics (*18–20*). In fact, almost all envisioned (and to date demonstrated) applications of VO$_2$ involve two-terminal devices that are driven electrically. However, despite its importance, very little is understood about the mechanisms underlying E-IMT. Crucially, the transformation pathway from the insulating monoclinic (*M1*) to metallic rutile (*R*) phase under electric-field remains unknown. In general, understanding the structural processes mediating electric-field-driven phase transitions remains a grand challenge in condensed matter physics. A major roadblock in addressing these issues has been the lack of a direct structural probe of the electrically-triggered transient state.

In contrast, there is a rich history of fundamental studies probing the ultrafast *photoinduced* IMT (P-IMT) (*5–7, 9*). Several spectroscopic and structural techniques have shown that femtosecond optical pulses trigger the transformation of *M1* to *R* on a picosecond timescale(*5–7*). In some cases, the structural and electronic transitions have been observed to be decoupled, pointing towards an isostructural, metallic monoclinic (*mM*) phase(*7, 8, 21*). Given that E-IMT and P-IMT occur on very different timescales (ns or μs versus ps or fs), it has generally been assumed that the two types of transformation follow distinct pathways, and connections between the two have not received much attention.

To further our understanding of electric-field effects and engineer the next generation of electronic switches based on correlated oxide semiconductors, it is essential to visualize atomic motions within the electrically-triggered transition state on fast timescales. Here, using a new stroboscopic mega-electron-volt (MeV) ultrafast electron diffraction technique, we report time-resolved measurements of structure following electrical excitation in VO$_2$ switches. By simultaneously probing changes in both structure and electronic transport under a pulsed electrical bias, we directly probe the mechanisms underlying E-IMT. These structural dynamics reveal a transient metallic monoclinic state on microsecond timescales, i.e. an isostructural phase that mediates the non-equilibrium electrically-triggered phase transition. In conjunction with phase-field simulations, we show that this metastable phase is locally stabilized by nanoscale disorder in the material. By exciting the devices with femtosecond optical pulses, we also trigger and probe P-IMT on the picosecond timescale. A comparison between E-IMT and P-IMT reveals a striking similarity in the transient structural dynamics, including the formation of the isostructural phase, suggesting universal features in the transition pathway across > 8 orders of magnitude in timescale. In contrast to earlier approaches which required heterostructure engineering or extreme experimental conditions(*2, 7, 22*), our approach introduces a purely electrically driven route to the isostructural phase in a typical polycrystalline film. This pathway is readily accessible in a large variety of device architectures that are commonly investigated in almost all solid-state device technology platforms. Broadly, this demonstrates electrical excitation to be a new mechanism for uncovering hidden order in correlated electron systems.

Figures 1A,B show a schematic of the experiment built at the MeV ultrafast electron diffraction (MeV-UED) facility(*23, 24*) at SLAC National Accelerator Laboratory. VO$_2$ switches were fabricated on 50 nm thick electron-transparent silicon nitride membranes. A representative quasi-static current-voltage curve is shown in Fig. 1C showing threshold switching behavior. To probe transient structural changes following electrical excitation, we developed a stroboscopic electrical pump-UED probe technique (see Methods, Fig. S1). Figure 1D shows the resistance of the VO$_2$ channel ($R_{VO2}$) as a function of delay time after the step voltage is applied. These transient characteristics are repeatable over millions of cycles (endurance > 30 million at 180 Hz), which is a crucial requirement for stroboscopic measurements (Fig. 1E).

The diffraction data were analyzed by azimuthally integrating the polycrystalline diffraction pattern and computing changes in intensity relative to the unexcited *M1* phase (see Methods). Figure 2A displays a two-dimensional color map of the normalized differential intensity, $\delta I = \Delta I/I$, as a function of momentum transfer $Q$ and delay time. A lineout at a delay of 500 μs is shown in Fig. 2B where three peaks of interest are indexed, namely $(30\bar{2})$, $(31\bar{3})$, and $(220)$. To compare the structural dynamics during E-IMT with P-IMT, we excited the same device with ~100 fs optical pulses from a 1.55 eV laser. As seen in Fig. 2C, photoexcitation induces ultrafast structural dynamics on the picosecond timescale. Figure 2D plots a lineout at a delay of 5 ps. Remarkably, even though their timescales differ by nearly 8 orders of magnitude, we observe very similar transient structures during E-IMT and P-IMT.

To resolve the timescales of the structural transformation following electrical excitation, we plot time-dependent $\delta I_{30\bar{2}}$, $\delta I_{31\bar{3}}$, and $\delta I_{220}$ at different voltages $V = 4$, 4.8 and 5.6 V (Fig. 3A-C). Three distinct regimes are identified. (i) As the voltage is turned on, the intensities of $(30\bar{2})$ and $(31\bar{3})$ peaks decrease, while the intensity of the $(220)$ peak increases. As shown by structure factor calculations (Fig. S2A), $(30\bar{2})$ and $(31\bar{3})$ peaks are present in the *M1* but absent in the high-symmetry *R* phase. Therefore, $\delta I_{30\bar{2}}$ and $\delta I_{31\bar{3}}$ capture the *M1*→*R* structural phase transformation (SPT), i.e. the nucleation and growth of the metallic *R* phase under electrical excitation. As we show later, the dynamics of the $(220)$ peak encodes information about the non-equilibrium structural phenomena occurring during the E-IMT. (ii) After the voltage is turned off, the structure persists for a finite duration, likely due to hysteresis in the phase transition as the device cools and the associated barrier for the reverse *R*→*M1* transformation(*12*). (iii) The structure returns to the *M1* phase in a quasi-exponential manner within ~2 ms, consistent with timescales for lateral heat transport through the VO$_2$/SiN membrane into the Si substrate.

In Fig. 3D-F, we zoom in to the rising edge of the voltage pulse and probe structural dynamics and transport with higher time resolution (also see Fig. S3). A delay is observed in the structural response (*via* $\delta I_{30\bar{2}}$ and $\delta I_{31\bar{3}}$) which is temporally correlated with a delayed response in $R_{VO2}$. We do not find any evidence of the *R* phase within this 'incubation' state to within the experimental detection limits, strongly suggesting that the direct electric-field induced transformation is small and that thermal effects are dominant(*25*)(see Supplementary Notes). To understand the dynamics of $\delta I_{220}$, we first note that the $(220)$ peak is present in both equilibrium phases. Structure factor calculations predict that the intensity of the $(220)$ peak in the *R* phase is higher than in the *M1* phase (Fig. S2A). To investigate whether this equilibrium transformation completely describes the observed positive changes in $\delta I_{220}$, we performed static electron diffraction measurements while heating the sample slowly (Fig. S4A). A broad transition beginning at 340 K was observed, with a width of ~20 K. In Fig. S4B we plot the temperature-dependent ratios of the differential intensity changes $\delta I_{220}/|\delta I_{30\bar{2}}|$ and $\delta I_{220}/|\delta I_{31\bar{3}}|$. Evaluating these at a temperature well above the transition temperature $T_c$, where the entire sample must transform to the R phase enables us to quantify the maximum relative change in $(220)$ intensity that can be caused by the equilibrium *M1*→*R* transition. In Fig. 3G we calculate these relative intensity changes for the time-resolved E-IMT and compare them with the equilibrium limits which are indicated by the dashed lines. Remarkably, at time delays smaller than 100 μs, we discover that the structural transformation under electrical excitation cannot be described solely by the formation of the *R* phase. At longer times, the relative peak changes tend towards their equilibrium values suggesting eventual completion of the *M1*→*R* SPT. Additional analysis is presented in Fig. S5.

To gain further insight into the unusual behavior of the (220) peak, we turn to the optical pump experiments. As shown in Fig. 3H, photoexcitation triggers the ultrafast $M1{\rightarrow}R$ SPT within ~0.5 – 0.7 ps, as indicated by a quenching of $(30\bar{2})$ and $(31\bar{3})$ peaks. In striking contrast, the (220) peak intensity increases on a significantly slower timescale of ~2.9 ps. Structure factor calculations (Fig. S2B,C) confirm that a simple model which considers a linear transformation of atom coordinates from $M1$ to $R$ cannot explain the ~4× difference in time constants(*26*). Previous optical pump-UED probe experiments have shown that the slow response of $\delta I_{220}$ is due to a symmetry-preserving orbital-selective charge reorganization in the $M1$ phase(*7*). This purely electronic transition results in a metastable metal-like monoclinic state (*mM*), which coexists with the thermodynamically stable $R$ phase. Recent optical pump-THz probe measurements(*8*) have supported this interpretation, as they report two distinct timescales in the THz conductivity response that closely match those measured in the UED structural dynamics. Our observation of a fast decay in $\delta I_{30\bar{2}}$ and $\delta I_{31\bar{3}}$, and a slow rise in $\delta I_{220}$ are in good agreement with these reports, demonstrating that photoexcitation creates a mixture of *mM* and *R* phases in our films. This is further validated by performing the same analysis as for E-IMT, where we compute the relative peak intensity changes with respect to the thermally-driven SPT. As shown in the inset of Fig. 3H, the effects are larger than can be explained simply by the $M1{\rightarrow}R$ transition. This analysis is consistent with the earlier experiments where it was seen that the ratio $\delta I_{220}/|\delta I_{30\bar{2}}|$ decreases with increasing pump excitation fluence; at intermediate fluences, a mixture of *mM*+*R* formed, whereas at high fluence, the sample converted fully to *R*(*8*). Taken together, our measurements provide clear evidence for the creation of a *mM* phase on ps timescales in photoexcited $VO_2$.

Returning to the analysis presented in Fig. 3G, this interpretation of P-IMT dynamics reveals an important discovery: Electrical excitation creates, on μs timescales, a transient *mM* phase in addition to the stable $R$ phase. This causes the peak intensity ratios $\delta I_{220}/|\delta I_{30\bar{2}}|$ and $\delta I_{220}/|\delta I_{31\bar{3}}|$ to exceed their equilibrium values. As the voltage is maintained, the *mM* domains convert fully to the thermodynamically stable $R$ phase within ~100 μs. To the best of our knowledge, this represents the first direct observation of this transient isostructural state during the electrically-triggered IMT in $VO_2$. Furthermore, this similarity between the pathways of E-IMT and P-IMT involving the intervening *mM* phase, is exemplified by the close correspondence between their structural dynamics across 8 orders of magnitude in timescale (Fig. 2).

To explain the transient emergence of the *mM* phase under electrical excitation, we perform time-dependent phase-field simulations of a multidomain $VO_2$ film under electric field(*27, 28*) (Fig. 4A, Methods). Based on temperature-dependent diffraction measurements which show a broad transition (Fig. S4A), we assume a spatially-inhomogeneous $T_c$ (Fig. 4B). This is modeled as a spatially-correlated random field with $\Delta T_c$ (i.e. the distribution in transition temperature) ranging 20 K and a correlation length of 25 nm (Fig. S6). In a polycrystalline film grown on a non-lattice matched substrate, this could be caused by subtle variations in the oxygen stoichiometry, strain, or other disorder associated with grain boundaries(*29, 30*). The nucleation and growth of new domains during the field-induced phase transition is described by the temporal and spatial evolution of the structural and electronic order parameters, free carrier density and temperature. This is done by minimizing the Landau free energy functional with respect to the order parameters while self-consistently solving the charge and heat diffusion equations (see Methods). Under the action of an electric-field applied to the $M1$ phase at $t = 0$, following an incubation period, domains with lower $T_c$ begin to nucleate the $R$ phase (Fig. 4C, Movie S1, Fig. S7). As this evolves, we find that local regions of the metastable *mM* phase form in order to minimize the overall interfacial energy of the system when these domains become smaller than a critical length scale (~10 nm). This mechanism is consistent with previous near-equilibrium studies in epitaxially-grown heterostructures wherein the *mM* phase, which is metastable in bulk, was stabilized through interfacial interactions with the *R* phase(*2*). The simulations additionally show that the intrinsic timescale for the formation of a single *mM* domain (white regions in Movie S1) could be smaller than 1 μs, suggesting that the 10-100 μs lifetime measured in the experiment is due to integration of the UED signal over the device area.

In two-terminal devices made of $VO_2$ or similar materials, it has been suggested that the E-IMT turn-on time is limited by electrical and thermal parasitics(*16*). While switching times of 0.5 – 10 ns have been demonstrated, the *intrinsic* speed limits of material transformation under electrical bias have been unclear(*14, 20, 25*). Our observation of similar transient phase dynamics during ultrafast P-IMT and slower E-IMT points toward an apparent universality in their transformation pathways, suggesting the potential for electrical turn-on time approaching a few ps in carefully-designed devices. Furthermore, given that the isostructural *mM* phase exists on μs or shorter timescales, it is likely that neuromorphic Mott oscillators operating at MHz and higher frequencies sample a complex phase space of structural and electronic states(*17, 31*). Deterministic engineering of disorder in nanoscale devices could open new avenues for the exploration of transient non-linear functions in strongly correlated materials for applications ranging from sensing to photonics. More generally, our results represent the first direct observation of an electrically-driven transient metastable phase in a solid-state device, and demonstrate a new method for inducing non-equilibrium phase transitions in much the same way as ultrafast photoexcitation has been used to unravel hidden order in materials(*32*). Our results will stimulate greater synergy between fundamental studies of ultrafast IMT and engineering efforts that leverage the unique switching functionalities for device applications, and these techniques can be broadly applied in the future to other phase changing systems.

**References**

1. Y. Wang, K. M. Kang, M. Kim, H. S. Lee, R. Waser, D. Wouters, R. Dittmann, J. J. Yang, H. H. Park, Mott-transition-based RRAM. *Mater. Today*. **28**, 63–80 (2019).

2. D. Lee, B. Chung, Y. Shi, G. Y. Kim, N. Campbell, F. Xue, K. Song, S. Y. Choi, J. P. Podkaminer, T. H. Kim, P. J. Ryan, J. W. Kim, T. R. Paudel, J. H. Kang, J. W. Spinuzzi, D. A. Tenne, E. Y. Tsymbal, M. S. Rzchowski, L. Q. Chen, J. Lee, C. B. Eom, Isostructural metal-insulator transition in VO2. *Science*. **362**, 1037–1040 (2018).

3. J. Laverock, S. Kittiwatanakul, A. Zakharov, Y. Niu, B. Chen, S. A. Wolf, J. W. Lu, K. E. Smith, Direct observation of decoupled structural and electronic transitions and an ambient pressure monocliniclike metallic phase of VO2. *Phys. Rev. Lett.* **113**, 216402 (2014).

4. M. M. Qazilbash, M. Brehm, B.-G. Chae, P.-C. Ho, G. O. Andreev, B.-J. Kim, S. J. Yun, A. V Balatsky, M. B. Maple, F. Keilmann, H.-T. Kim, D. N. Basov, Mott transition in VO2 revealed by infrared spectroscopy and nano-imaging. *Science*. **318**, 1750–1753 (2007).

5. A. Cavalleri, C. Tóth, C. W. Siders, J. A. Squier, F. Ráksi, P. Forget, J. C. Kieffer, Femtosecond structural dynamics in VO2 during an ultrafast solid-solid phase transition.

*Phys. Rev. Lett.* **87**, 237401 (2001).

6. P. Baum, D. S. Yang, A. H. Zewail, 4D visualization of transitional structures in phase transformations by electron diffraction. *Science*. **318**, 788–792 (2007).

7. V. R. Morrison, R. P. Chatelain, K. L. Tiwari, A. Hendaoui, A. Bruhács, M. Chaker, B. J. Siwick, A photoinduced metal-like phase of monoclinic VO2 revealed by ultrafast electron diffraction. *Science*. **346**, 445–448 (2014).

8. M. R. Otto, L. P. R. de Cotret, D. A. Valverde-Chavez, K. L. Tiwari, N. Émond, M. Chaker, D. G. Cooke, B. J. Siwick, How optical excitation controls the structure and properties of vanadium dioxide. *Proc. Natl. Acad. Sci. U. S. A*. **116**, 450–455 (2018).

9. S. Wall, S. Yang, L. Vidas, M. Chollet, J. M. Glownia, M. Kozina, T. Katayama, T. Henighan, M. Jiang, T. A. Miller, D. A. Reis, L. A. Boatner, O. Delaire, M. Trigo, Ultrafast disordering of vanadium dimers in photoexcited VO2. *Science*. **362**, 572–576 (2018).

10. M. Liu, H. Y. Hwang, H. Tao, A. C. Strikwerda, K. Fan, G. R. Keiser, A. J. Sternbach, K. G. West, S. Kittiwatanakul, J. Lu, S. A. Wolf, F. G. Omenetto, X. Zhang, K. A. Nelson, R. D. Averitt, Terahertz-field-induced insulator-to-metal transition in vanadium dioxide metamaterial. *Nature*. **487**, 345–348 (2012).

11. A. X. Gray, M. C. Hoffmann, J. Jeong, N. P. Aetukuri, D. Zhu, H. Y. Hwang, N. C. Brandt, H. Wen, A. J. Sternbach, S. Bonetti, A. H. Reid, R. Kukreja, C. Graves, T. Wang, P. Granitzka, Z. Chen, D. J. Higley, T. Chase, E. Jal, E. Abreu, M. K. Liu, T. C. Weng, D. Sokaras, D. Nordlund, M. Chollet, R. Alonso-Mori, H. Lemke, J. M. Glownia, M. Trigo, Y. Zhu, H. Ohldag, J. W. Freeland, M. G. Samant, J. Berakdar, R. D. Averitt, K. A. Nelson, S. S. P. Parkin, H. A. Dürr, Ultrafast terahertz field control of electronic and structural interactions in vanadium dioxide. *Phys. Rev. B*. **98**, 045104 (2018).

12. J. del Valle, P. Salev, F. Tesler, N. M. Vargas, Y. Kalcheim, P. Wang, J. Trastoy, M. H. Lee, G. Kassabian, J. G. Ramírez, M. J. Rozenberg, I. K. Schuller, Subthreshold firing in Mott nanodevices. *Nature*. **569**, 388–392 (2019).

13. G. Stefanovich, A. Pergament, D. Stefanovich, Electrical switching and Mott transition in VO2. *J. Phys. Condens. Matter*. **12**, 8837–8845 (2000).

14. Y. Zhou, X. Chen, C. Ko, Z. Yang, C. Mouli, S. Ramanathan, Voltage-triggered ultrafast phase transition in vanadium dioxide switches. *IEEE Electron Device Lett.* **34**, 220–222 (2013).

15. N. Shukla, A. V. Thathachary, A. Agrawal, H. Paik, A. Aziz, D. G. Schlom, S. K. Gupta, R. Engel-Herbert, S. Datta, A steep-slope transistor based on abrupt electronic phase transition. *Nat. Commun.* **6**, 7812 (2015).

16. Y. Zhou, S. Ramanathan, Mott Memory and Neuromorphic Devices. *Proc. IEEE*. **103**, 1289–1310 (2015).

17. W. Yi, K. K. Tsang, S. K. Lam, X. Bai, J. A. Crowell, E. A. Flores, Biological plausibility and stochasticity in scalable VO2 active memristor neurons. *Nat. Commun.* **9**, 4661 (2018).

18. T. Driscoll, H.-T. Kim, B.-G. Chae, B.-J. Kim, Y.-W. Lee, N. M. Jokerst, S. Palit, D. R.

Smith, M. Di Ventra, D. N. Basov, Memory Metamaterials. *Science*. **325**, 1518–1522 (2009).

19. L. Liu, L. Kang, T. S. Mayer, D. H. Werner, Hybrid metamaterials for electrically triggered multifunctional control. *Nat. Commun.* **7**, 13236 (2016).

20. P. Markov, R. E. Marvel, H. J. Conley, K. J. Miller, R. F. Haglund, S. M. Weiss, Optically Monitored Electrical Switching in VO2. *ACS Photonics*. **2**, 1175–1182 (2015).

21. D. Wegkamp, M. Herzog, L. Xian, M. Gatti, P. Cudazzo, C. L. McGahan, R. E. Marvel, R. F. Haglund, A. Rubio, M. Wolf, J. Stähler, Instantaneous band gap collapse in photoexcited monoclinic VO2 due to photocarrier doping. *Phys. Rev. Lett.* **113**, 216401 (2014).

22. E. Arcangeletti, L. Baldassarre, D. Di Castro, S. Lupi, L. Malavasi, C. Marini, A. Perucchi, P. Postorino, Evidence of a pessure-induced metallization process in monoclinic VO2. *Phys. Rev. Lett.* **98**, 196406 (2007).

23. S. P. Weathersby, G. Brown, M. Centurion, T. F. Chase, R. Coffee, J. Corbett, J. P. Eichner, J. C. Frisch, A. R. Fry, M. Gühr, N. Hartmann, C. Hast, R. Hettel, R. K. Jobe, E. N. Jongewaard, J. R. Lewandowski, R. K. Li, A. M. Lindenberg, I. Makasyuk, J. E. May, D. McCormick, M. N. Nguyen, A. H. Reid, X. Shen, K. Sokolowski-Tinten, T. Vecchione, S. L. Vetter, J. Wu, J. Yang, H. A. Dürr, X. J. Wang, Mega-electron-volt ultrafast electron diffraction at SLAC National Accelerator Laboratory. *Rev. Sci. Instrum.* **86**, 073702 (2015).

24. X. Shen, R. K. Li, U. Lundström, T. J. Lane, A. H. Reid, S. P. Weathersby, X. J. Wang, Femtosecond mega-electron-volt electron microdiffraction. *Ultramicroscopy*. **184**, 172–176 (2018).

25. J. S. Brockman, L. Gao, B. Hughes, C. T. Rettner, M. G. Samant, K. P. Roche, S. S. P. Parkin, Subnanosecond incubation times for electric-field-induced metallization of a correlated electron oxide. *Nat. Nanotechnol.* **9**, 453–458 (2014).

26. Z. Tao, F. Zhou, T. R. T. Han, D. Torres, T. Wang, N. Sepulveda, K. Chang, M. Young, R. R. Lunt, C. Y. Ruan, The nature of photoinduced phase transition and metastable states in vanadium dioxide. *Sci. Rep.* **6**, 38514 (2016).

27. Y. Shi, L. Q. Chen, Phase-field model of insulator-to-metal transition in VO2 under an electric field. *Phys. Rev. Mater.* **2**, 053803 (2018).

28. Y. Shi, L. Q. Chen, Current-driven insulator-to-metal transition in strongly correlated VO2. *Phys. Rev. Appl.* **11**, 014059 (2019).

29. J. Nag, R. F. Haglund, Synthesis of vanadium dioxide thin films and nanoparticles. *J. Phys. Condens. Matter*. **20**, 264016 (2008).

30. T. J. Huffman, D. J. Lahneman, S. L. Wang, T. Slusar, B. J. Kim, H. T. Kim, M. M. Qazilbash, Highly repeatable nanoscale phase coexistence in vanadium dioxide films. *Phys. Rev. B*. **97**, 085146 (2018).

31. N. Shukla, A. Parihar, E. Freeman, H. Paik, G. Stone, V. Narayanan, H. Wen, Z. Cai, V. Gopalan, R. Engel-Herbert, D. G. Schlom, A. Raychowdhury, S. Datta, Synchronized charge oscillations in correlated electron systems. *Sci. Rep.* **4**, 4964 (2014).

32. D. N. Basov, R. D. Averitt, D. Hsieh, Towards properties on demand in quantum materials. *Nat. Mater.* **16**, 1077–1088 (2017).

33. S. M. Bohaichuk, S. Kumar, G. Pitner, C. J. McClellan, J. Jeong, M. G. Samant, H. S. P. Wong, S. S. P. Parkin, R. S. Williams, E. Pop, Fast spiking of a Mott VO2-carbon nanotube composite device. *Nano Lett.* **19**, 6751–6755 (2019).

34. V. Eyert, The metal-insulator transitions of VO2: A band theoretical approach. *Ann. der Phys.* **11**, 650–702 (2002).

35. L. M. Peng, G. Ren, S. L. Dudarev, M. J. Whelan, Robust parameterization of elastic and absorptive electron atomic scattering factors. *Acta Crystallogr.* **A52**, 257–276 (1996).

36. C. R. Everhart, J. B. MacChesney, Anisotropy in the electrical resistivity of vanadium dioxide single crystals. *J. Appl. Phys.* **39**, 2872–2874 (1968).

37. K. Sokolowski-Tinten, X. Shen, Q. Zheng, T. Chase, R. Coffee, M. Jerman, R. K. Li, M. Ligges, I. Makasyuk, M. Mo, A. H. Reid, B. Rethfeld, T. Vecchione, S. P. Weathersby, H. A. Dürr, X. J. Wang, Electron-lattice energy relaxation in laser-excited thin-film Au-insulator heterostructures studied by ultrafast MeV electron diffraction. *Struct. Dyn.* **4**, 054501 (2017).

**Acknowledgments**

This work is supported primarily by the U.S. Department of Energy, Office of Science, Basic Energy Sciences, Materials Sciences and Engineering Division, under Contract DE-AC02-76SF00515. MeV-UED is operated as part of the Linac Coherent Light Source at the SLAC National Accelerator Laboratory, supported by the U.S. Department of Energy, Office of Science, Office of Basic Energy Sciences under Contract No. DE-AC02-76SF00515. Work was performed in part at the Stanford Nanofabrication Facility and the Stanford Nano Shared facilities which receive funding from the National Science Foundation as part of the National Nanotechnology Coordinated Infrastructure Award ECCS-1542152. S.R. acknowledges AFOSR FA9550-18-1-0250 for support. The effort of Y.S. and L.Q.C. is supported as part of the Computational Materials Sciences Program funded by the U.S. Department of Energy, Office of Science, Basic Energy Sciences, under Award Number DE-SC0020145. A.S. acknowledges support from the U.S. Department of Energy, Office of Science, Basic Energy Sciences under award number DE-SC-0012375. The authors thank Philipp Muscher, Alexander Reid, Stephen Weathersby and Stephanie Bohaichuk for helpful discussions, and Qiyang Lu, Allen Liang and Andrey Poletayev for a critical reading of the manuscript.

**Author contributions**

A.S. and A.M.L. conceived the project; A.M.L., W.C., X.W. and S.R. supervised the experiments; L.Q.C. supervised the theoretical calculations; A.S. fabricated the devices with contributions from S.K., M.Z. and S.J.P.; Y.S. and S.R. grew VO$_2$ films; A.S. led the design of the electrical-pump setup and performed UED experiments with X.S., S.J.P., and M.Z.; A.S., X.S. and A.M.L analyzed and interpreted the data; Y.S. and L.Q.C. performed phase-field simulations; A.S. wrote the manuscript with inputs from all authors; A.M.L. directed the overall research.

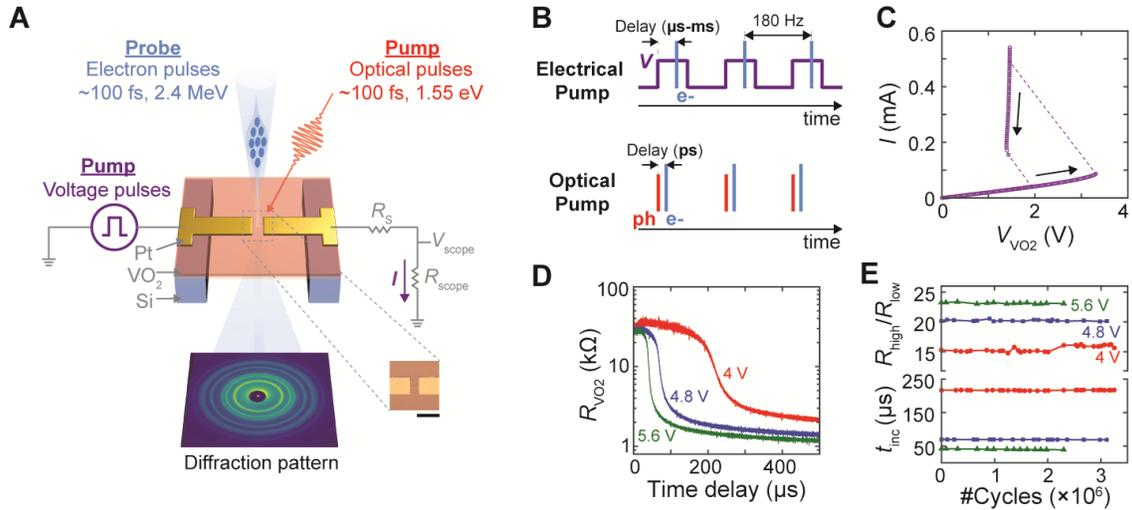

**Fig. 1. Stroboscopic pump-probe setup.** (A) Schematic of the experimental setup showing the $VO_2$ membrane device excited by voltage and laser pulses and probed by MeV electron pulses in transmission mode. Scale bar in the zoomed-in optical micrograph is 50 μm. (B) Stroboscopic measurement scheme showing voltage and laser pulses synchronized to the 180 Hz electron probe pulses with tunable delay. (C) Representative steady-state characteristics under voltage control. (D) Transient electrical characteristics showing device resistance as a function of time after a step voltage is applied. (E) Stability of the resistance switching ratio and incubation time over millions of cycles.

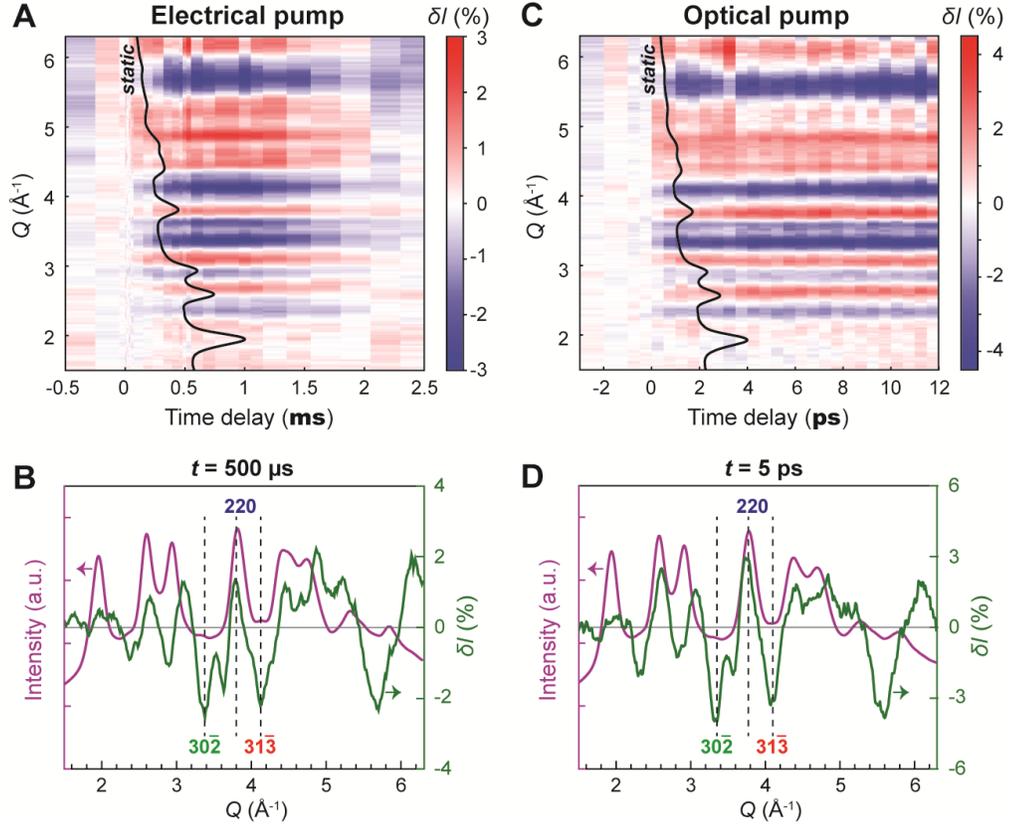

**Fig. 2. Similarity of transient structures formed under electrical and optical excitation.** (A) Differential intensity changes during E-IMT with a 5.6 V, 500 μs pulse. The static diffraction pattern is shown by the black curve. (B) Lineout at a delay of 500 μs showing the static pattern (purple; intensity is multiplied by $Q^{1.4}$ to aid visualization) and the differential intensity change (green). (C) Differential intensity changes during P-IMT with a ~100 fs pulse at a fluence of ~49 mJ cm$^{-2}$. (D) Lineout at a delay of 5 ps. In (B) and (D) three important peaks are indexed: $(30\bar{2})$, $(31\bar{3})$, and $(220)$.

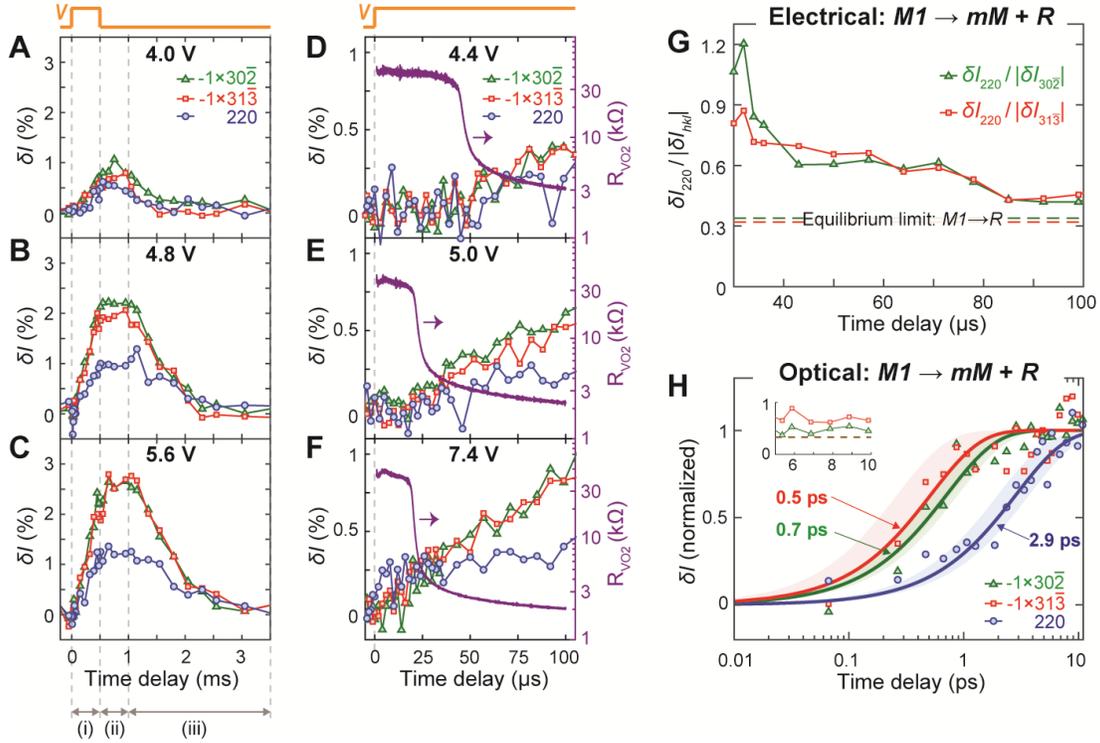

**Fig. 3. Phase transformation dynamics and evidence of a transient metallic monoclinic (*mM*) phase.** (A)-(C) Structural dynamics during and following electrical excitation with a 500 µs pulse of magnitude 4, 4.8 and 5.6 V on Device 1 (width = 40 µm, length = 20 µm). (D)-(F) Short timescale structural dynamics following a step voltage applied at $t$ = 0 of magnitude 4.4, 5 and 7.4 V on Device 2 (width = length = 20 µm). The transient device resistance $R_{VO2}$ is measured simultaneously (purple). (G) Ratio of differential peak intensity changes of the $(220)$ and $(30\bar{2})$ peaks (green triangles), and $(220)$ and $(31\bar{3})$ peaks (red squares) corresponding to the data shown in (F). Green and red dashed lines indicate the equilibrium limits for *M1→R* SPT. (H) Normalized differential intensity change following photoexcitation with ~100 fs optical pulses at a fluence ~27 mJ cm$^{-2}$. Solid lines are single exponential fits and the shaded areas represent 95 % confidence intervals. The inset shows the ratio of differential intensity changes $\delta I_{220}/|\delta I_{30\bar{2}}|$ and $\delta I_{220}/|\delta I_{31\bar{3}}|$, compared to the equilibrium limits indicated by the dashed lines. Data in (G) and the inset of (H) have been smoothed by 3-point adjacent averaging for better visualization.

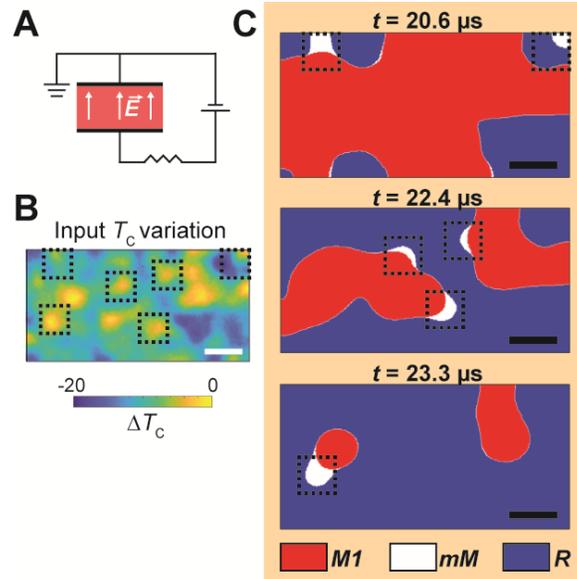

**Fig. 4. Phase-field simulations predict a transient *mM* phase following electrical excitation.** (A) Schematic showing a 2D VO$_2$ device under an in-plane electric-field. (B) The multidomain film is assumed to have a spatially heterogeneous variation of transition temperature ($T_c$) due to nanoscale disorder. (C) Phase maps at different time delays after an electric-field of magnitude 3.7 kV cm$^{-1}$ is applied at $t = 0$. Interfacial interactions between the equilibrium *M1* and *R* phases stabilize the *mM* phase on a μs timescale (see regions within the dashed boxes). Scale bars in (B) and (C) are 20 nm.

**Methods**

Stroboscopic electrical-pump ultrafast electron diffraction (UED)-probe setup

The devices consist of 60 nm thick polycrystalline $VO_2$ films grown by physical vapor deposition on 50 nm thick SiN membranes. Metal electrodes (3 nm Ti/50 nm Pt) were patterned photolithographically to create two-terminal devices. Details of the fabrication process are given below. Two devices were measured at UED. Device 1 had a channel width of 40 µm and a length of 20 µm (data in Fig. 1D,E, Fig. 2, Fig. 3A-C). Device 2 had a channel width of 20 µm and a length of 20 µm (data in Fig. 3D-F, Fig. S3, Fig. S5). Channel length is defined as the separation between the electrodes. The device was excited by a periodic train of voltage pulses with amplitude higher than the threshold voltage, and pulse widths ranging from 100 to 500 µs. These were produced by a pulse delay generator that was synchronized to the MeV electron pulses at a repetition rate of 180 Hz. Details of the electrical setup and timing scheme are presented below and in Fig. S1. Each voltage pulse triggered a transformation from the insulating to metallic state, which was confirmed by measuring the resistance of the device. The material reverted to the insulating state after the pulse was turned off. Time-resolved changes in the structure were measured by diffraction using a periodic train of ~100 fs, ~2.4 MeV electron pulses by varying the delay between the applied voltage 'pump' and electron 'probe' pulses. The electron beam spot size was approximately 70 µm full width at half maximum (FWHM). The resistance of the device was simultaneously monitored by measuring the current in the circuit using a high-speed oscilloscope connected in series. A large series resistor ($R_s$ = 5 kΩ) was employed to limit the current in the high conductance state and improve switching endurance(33).

Device fabrication

We started with ~300 µm thick Si wafers (p-type, 1-30 Ohm-cm, <100> orientation) that were coated on both sides with 50 nm thick low-stress SiN grown by low pressure chemical vapor deposition. Windows were opened in SiN on the bottom-side using standard photolithography and dry etching techniques. The silicon substrate was back-etched using a 30 % KOH solution at 85°C for ~6 hours, resulting in free-standing SiN membranes (~200 × 200 µm, 50 nm thick). The etched wafers were decontaminated by immersing in a 5:1:1 $H_2O:H_2O_2:HCl$ solution for ~30 minutes and rinsing in de-ionized water. 60 nm thick polycrystalline $VO_2$ films were grown on the top-side using physical vapor deposition. The background pressure of the growth chamber was < $1 \times 10^{-7}$ Torr before deposition. $V_2O_5$ was used as a target equipped with a radio frequency (RF) power supply. For a specific growth, the growth background pressure was set to $5 \times 10^{-3}$ Torr, and the atmosphere was a mixture of 49.6 standard cubic centimeters per minute (sccm) Ar and 0.4 sccm $O_2$. The growth temperature was fixed at 700°C and the target power was 100 W (growth rate ~36 nm per hour). The films were annealed in an $O_2$ atmosphere at 400°C for 50 minutes and 1500 mTorr to enhance the switching performance. Metal electrodes (3 nm Ti/50 nm Pt) were patterned on the top-side using photolithography and deposited *via* electron-beam evaporation. After initial electrical testing in a probe station, the devices were mounted onto a chip carrier and wire-bonded for UED measurements.

Electrical connections

The devices were wire bonded to chip carriers, which were soldered to ultrahigh vacuum (UHV) compatible BNC cables (see Fig. S1). The current-limiting resistor was soldered to the chip carrier on the ground side. Feedthroughs in the sample chamber were used to connect the devices and reference photodiode to biasing and sensing electronics placed outside. All measurements were performed under vacuum (pressure ~$1 \times 10^{-6}$ Torr) at a baseline temperature of 300 K. Fig. 1A and Fig. S1B show the electrical circuit used to induce and probe E-IMT. Electrical pulses were applied using a pulse delay generator (Stanford Research Systems, SRS DG645). The generator was triggered externally using a 180 Hz reference signal to synchronize the applied bias pulses with the MeV electron probe pulses. The electrical state of the $VO_2$ was monitored in real time by measuring the current through the circuit. This was done by probing the voltage across the 50 Ω input of a 6 GHz oscilloscope (Keysight, DSO 90604A) placed in series with the device.

Timing

Relative timing between the 1.55 eV optical and MeV electron pulses was determined by monitoring the diffraction pattern of a single crystal (100) oriented gold film. We monitored a decrease in the intensity of Bragg peaks upon photoexcitation due to the Debye Waller effect. This enabled determination of time-zero between the optical and electron probe pulses to within <1 ps. To determine relative timing between the applied voltage bias and MeV electron pulses, we mounted a fast photodiode next to the devices. We measured the arrival time of the 1.55 eV optical pulses on this diode while simultaneously recording the voltage pulses applied across the device on the oscilloscope. Since the optical and MeV electron pulses are synchronized in time, time-zero was determined between the voltage pump and MeV probe pulses.

UED data analysis

The diffraction pattern on the phosphor screen was imaged by a charge coupled device (CCD) camera. The CCD background was removed by subtracting the average counts at the corners of each frame. Images were corrected for ellipticity and circles were fit to diffraction rings in order to locate the centroid of the pattern. Azimuthal integration was performed and the data were normalized to the total intensity between $Q$ = 1.5 to 6 Å$^{-1}$ for data shown in Fig. 2, Fig. 3A-C, and between $Q$ = 1.5 to 9 Å$^{-1}$ for data shown in Fig. 3D-H, Fig. S4, Fig. S5. The larger $Q$ range in the latter data set was due to a shorter sample-to-detector distance used in those experiments. For each set of measurements, reciprocal lattice distances were calibrated by taking the static diffraction pattern of a single crystal (100) oriented gold film.

$I(Q)$ at a given time delay was calculated by averaging images taken over multiple scans. The delay times, in both optical and electrical pump experiments were varied in random order. Differential intensity changes, $\delta I(Q, t) = [I(Q, t) - I_0(Q)]/I_0(Q)$, were calculated by comparing $I(Q)$ of the excited state to that of the unexcited state $I_0(Q)$. To obtain $\delta I_{hkl}(t)$ for a specific Bragg peak, we averaged $\delta I(Q, t)$ in a ± 0.04 Å$^{-1}$ window around the peak. This window size was found to be optimal as it was large enough to average enough data for good signal-to-noise, while not being so large as to include adjacent peaks.

Transient electrical measurements and analysis

We measured the transient electrical resistance after application of a step voltage $V_{app}$ using the circuit shown in Fig. 1A and Fig. S1B. Current flowing through the device was measured by connecting a high-speed oscilloscope in series and monitoring the voltage $V_{osc}$ dropped across its 50 Ω input. This is given by $i(t) = V_{osc}(t)/50$. Due to the large capacitance of the device, which is likely dominated by the metal contacts, we observed a spike in $i(t)$ that lasts ~10 µs after $V_{app}$ is applied, as shown in Fig. S8. We interpret this as coming from the characteristic charging time of the capacitor formed by the metal electrodes, $C_{el}$. The total current is the sum of this capacitive component $i_{C_{el}}$, and a resistive component that flows through the VO₂ film $i_{R_{VO2}}$, i.e. $i(t) = i_{C_{el}}(t) + i_{R_{VO2}}(t)$. For short timescales ($t < t_{inc}$, the incubation time) where the current spike is observed, we can assume $i_{R_{VO2}}$ to be a constant value coming from the finite resistivity of the insulating *M1* state. The capacitor charging current can be modeled with an expression of the form: $i_{C_{el}}(t) = i_{C_{el},0}\exp(-t/\tau)$, where $\tau$ is the time constant. To remove the capacitive component, we fitted $i(t)$ to an equation of the form $i(t) = i_{C_{el},0}\exp(-t/\tau) + i_{R_{VO2},0}$ for $0.5 < t < 10$ µs, and extracted $i_{C_{el},0}, i_{R_{VO2},0}$ and $\tau$. We measured $\tau \sim 2.5 - 2.7$ and $2.2 - 2.6$ µs for Device 1 and 2, respectively, across all applied voltages $V_{app}$. The current flowing through the VO₂ film is given by, $i_{R_{VO2}}(t) = i(t) - i_{C_{el},0}\exp(-t/\tau)$.

The resistance of the VO₂ channel was calculated by subtracting the voltage dropped across the series resistor $R_s$ and the 50 Ω oscilloscope resistance,

$$R_{VO2}(t) = \frac{V_{app} - i_{R_{VO2}}(t)[R_s + 50]}{i_{R_{VO2}}(t)} \tag{S1}$$

The oscilloscope samples data with an interval of 50 – 200 ps. To improve the signal-to-noise ratio of $R_{VO2}(t)$ we smoothed the data by adjacent averaging within a 10 ns window. This is justified because the timescales of the electrical transients are > 1 µs. To quantify the stability of cycling, we define the 'high' and 'low' state resistances as $R_{VO2}$ at $t = 15$ and 450 µs, respectively, for Device 1 data in Fig. 1D and at $t = 10$ and 95 µs, respectively, for Device 2 data in Fig. S3A. The incubation time was defined as $t$ at which $R_{VO2} = 10$ kΩ. The resistance switching ratio $R_{high}/R_{low}$ and $t_{inc}$ do not show significant variation with cycle number over the duration of the UED measurement (see Fig. 1E and Fig. S3B). For Device 2, at one of the voltages (7.4 V), we observe a decrease in the switching ratio and incubation time with cycling (Fig. S3B). However, as shown by the analysis in Fig. S5D-F, our main conclusions are robust to this variation.

Structure factor calculations

To calculate the expected diffraction intensities of the equilibrium *M1* and *R* phases, we start by calculating the structure factor $S_{hkl}(Q)$ for the set of lattice planes $(hkl)$,

$$S_{hkl}(Q) = \sum_j f_j(Q)\cdot\exp[-i2\pi(hx_j + ky_j + lz_j)]. \tag{S2}$$

$Q = 2\pi/d_{hkl}$, where $d_{hkl}$ is the d-spacing between planes belonging to the (hkl) family. $f_j(Q)$ is the atomic scattering factor of the jth atom in the unit cell, and $(x_j, y_j, z_j)$ are its fractional coordinates taken from literature(34). $f_j(Q)$ is calculated using the following equation,

$$f_j(Q) = \sum_{i=1}^{5} \alpha_{j,i} \exp\left[-\beta_{j,i}\left(\frac{Q}{4\pi}\right)^2\right]. \tag{S3}$$

Fitting parameters $\alpha_{j,i}$ and $\beta_{j,i}$ for atom $j$ (here, $j$ = V or O) are taken from literature(35). Finally, the peak intensity is obtained, $I(Q) = |S(Q)|^2$. This is plotted in Fig. S2 for the equilibrium M1 and R structures. For the purposes of visualization, we have applied a Gaussian broadening ($\Delta Q_{FWHM}$ = 0.085 Å$^{-1}$) to each peak.

Phase-field modeling

Recently we have formulated a phase-field model of the IMT in VO$_2$ incorporating explicitly the structural and electronic instabilities and the effect of free carriers, with the assumption of isothermal conditions(27, 28). To better describe the conditions of the present experiment, we now include the Joule heating effect in the phase-field model. The evolution of the VO$_2$ system subject to a voltage can be characterized in terms of the partial equilibrium free energy $F[T(\boldsymbol{r},t), \Phi(\boldsymbol{r},t); \mu(\boldsymbol{r},t), \eta(\boldsymbol{r},t), n(\boldsymbol{r},t), p(\boldsymbol{r},t)]$, where $T(\boldsymbol{r},t)$ is the temperature field, $\Phi(\boldsymbol{r},t)$ the electric potential field, $\mu(\boldsymbol{r},\text{t})$ the electronic order parameter field, $\eta(\boldsymbol{r},t)$ the structural order parameter field, $n(\boldsymbol{r},t)$ the free electron density field, and $p(\boldsymbol{r},t)$ the free hole density field. $\boldsymbol{r}$ is the spatial coordinate and $t$ is time. The free electrons in the conduction band and the free holes in the valence band are assumed to be in partial equilibrium within themselves, i.e. the free electrons and free holes obey a Fermi distribution and have their own local quasi-chemical potentials. This partial equilibrium also implies a well-defined local temperature (which is the same for the lattice and electrons) and order parameters around a given spatial point. The system has not yet achieved full equilibrium among all these local degrees of freedom. $F$ is a functional of the fields,

$$F = \int [f_b(T;\mu,\eta) + f_g(\nabla\mu, \nabla\eta) + f_{eh}(T, \Phi; \mu, n, p)]d^3r, \tag{S4}$$

where $f_b$ is the bulk Landau-potential energy density, $f_g$ is the gradient energy density, and $f_{eh}$ is the free energy density of free electrons and holes. The detailed form of $f_b$ and $f_g$ and values of the Landau parameters in them can be found in Ref. (2). The detailed form of $f_{eh}$ and values of the parameters in it can be found in Ref. (28).

The evolution of the system is described by the Allen-Cahn equations for $\mu$ and $\eta$,

$$\frac{\partial \mu}{\partial t} = -L_1 \frac{\delta F}{\delta \mu}, \tag{S5a}$$

$$\frac{\partial \eta}{\partial t} = -L_2 \frac{\delta F}{\delta \eta}, \tag{S5b}$$

and the Cahn-Hilliard equations for $n$ and $p$,

$$\frac{\partial n}{\partial t} = \nabla \cdot \left(\frac{\mathbf{M}_e n}{e} \cdot \nabla \frac{\delta F}{\delta n}\right) + s, \tag{S6a}$$

$$\frac{\partial p}{\partial t} = \nabla \cdot \left(\frac{\mathbf{M}_h p}{e} \cdot \nabla \frac{\delta F}{\delta p}\right) + s, \tag{S6b}$$

together with the Poisson equation for $\Phi$,

$$-\nabla^2 \Phi = \frac{e(p-n)}{\epsilon_0 \epsilon}, \tag{S7}$$

and the heat equation for $T$,

$$\frac{\partial u}{\partial t} = \nabla \cdot (\kappa \nabla T) + \mathbf{J} \cdot \boldsymbol{\sigma}^{-1} \cdot \mathbf{J} + s_h. \tag{S8}$$

Here $L_1$ and $L_2$ are constants, $\mathbf{M}_e$ and $\mathbf{M}_h$ are the electron and hole mobility tensors, respectively, $e$ is the elementary charge (positive), $\epsilon_0$ is the vacuum permittivity, and $\epsilon$ is the relative permittivity. $s$ is a source term representing the electron-hole recombination (*28*). We explain the heat equation in more detail here. It is essentially an equation of energy conservation. $u$ is the internal energy density, which can be written in the form $u = C_v T + u_L$, where $C_v$ is the volumetric specific heat of the $R$ phase and $u_L$ is the internal energy density arising from $F$, since the free energy $F$ is referencing the high-temperature $R$ phase. For simplicity, we can neglect the electronic contribution to $u$, since the specific heat of the lattice is typically much larger than that of electrons near room temperature. Then $u_L$ is

$$u_L \approx f_b - T \frac{\partial f_b}{\partial T} = f_b(T=0; \mu, \eta), \tag{S9}$$

in which the second equality is due to the linear dependence of $f_b$ on $T$. The $u_L$ term in the heat equation accounts for absorption or release of the latent heat during the first-order IMT. The heat equation is finally

$$C_v \frac{\partial T}{\partial t} + \frac{\partial f_b(0; \mu, \eta)}{\partial \mu} \frac{\partial \mu}{\partial t} + \frac{\partial f_b(0; \mu, \eta)}{\partial \eta} \frac{\partial \eta}{\partial t} = \nabla \cdot (\kappa \nabla T) + \mathbf{J} \cdot \boldsymbol{\sigma}^{-1} \cdot \mathbf{J} + s_h \tag{S10}$$

where $\kappa$ is the thermal conductivity, $\mathbf{J}$ is the current density, and $\sigma = e(\mathbf{M}_e n + \mathbf{M}_h p)$ is the electrical conductivity tensor. In a 2D system abstracting the 3D VO$_2$/substrate system, the heat dissipation into the substrate can be characterized by the source term $s_h = h(T_s - T)/L_{z,VO2}$, where $h$ is the heat transfer coefficient, $T_s$ is the ambient temperature, and $L_{z,VO2}$ is the thickness of the VO$_2$ film in the direction perpendicular to the 2D plane. We estimate $h$ using a thermal circuit model of the actual device as described in the next section.

We consider a 2D rectangular system to abstract the real 3D system in the experiment yet adequate to capture the main experimental features (see Fig. 4A). The rectangular VO2 channel has a width $L_x$ in the $x$ direction (rutile $a$ axis) and a length $L_y$ in the $y$ direction (rutile $c$ axis). The mobility tensors can be considered diagonal $\mathbf{M}_e = \text{diag}(M_{ea}, M_{ec})$ and $\mathbf{M}_h = \text{diag}(M_{ha}, M_{hc})$ with $M_{ea} \sim 0.5 M_{ec}$, $M_{ha} \sim 0.5 M_{hc}$,(36) and $M_{hc} \sim M_{ec}/1.2$.(28) Because the VO2 sample in the experiment is not a perfect single crystal, we consider an effective electron mobility $M_{ec} \sim 0.0125$ cm$^2$V$^{-1}$s$^{-1}$ and an effective band gap of the $M1$ phase ~0.115 eV reproducing the conductivities of both the $M1$ phase (near $T_c$) and the $R$ phase in the experiment. The components of the current density can be written as $J_x = e(j_{hx} - j_{ex})$ and $J_y = e(j_{hy} - j_{ey})$; $j_{ex} = -M_{ea} n \partial_x \xi_e / e$, $j_{ey} = -M_{ec} n \partial_y \xi_e / e$, $j_{hx} = -M_{ha} p \partial_x \xi_h / e$, and $j_{hy} = -M_{hc} p \partial_y \xi_h / e$ are the components of the number current density of free electrons and free holes, respectively, where $\xi_e = \delta F / \delta n$ and $\xi_h = \delta F / \delta p$ are the quasi-chemical potentials of free electrons and free holes, respectively. The Joule heating power density is then $\mathbf{J} \cdot \boldsymbol{\sigma}^{-1} \cdot \mathbf{J} = e(j_{hx} - j_{ex})^2 / (M_{ea} n + M_{ha} p) + e(j_{hy} - j_{ey})^2 / (M_{ec} n + M_{hc} p)$.

The four boundaries lie at $y = 0$, $x = L_x$, $y = L_y$, and $x = 0$. The VO2 channel is connected in series with a resistor $R_s$ and a voltage source $V$ through metal electrodes. The voltage drop is in the $y$ direction. According to the physical setup, the boundary conditions are as follows. $\mu$ and $\eta$ have Neumann boundary conditions with zero flux at all boundaries representing no interaction between the order parameters and the surrounding. For $n$, $j_{ex}|_{x=0} = j_{ex}|_{x=L_x} = 0$, $\xi_e|_{y=0} - \xi_{eq} = -e\Phi|_{y=0}$, and $\xi_e|_{y=L_y} - \xi_{eq} = 0$. $\xi_{eq}$ is the equilibrium intrinsic chemical potential of free electrons (in the notation of the phase-field model, $-\xi_{eq}$ is the equilibrium intrinsic chemical potential of free holes, and in this specific case $\xi_{eq} = 0$)(28). The boundary conditions for $n$ at $x = 0$ and $x = L_x$ are straightforward. The boundary conditions for $n$ at $y = 0$ and $y = L_y$ reflect the essence of a voltage source, that is, the electromotive force is a manifestation of the electron chemical potential difference between the anode and cathode. For $p$, similar to the case of $n$, $j_{hx}|_{x=0} = j_{hx}|_{x=L_x} = 0$, $\xi_h|_{y=0} + \xi_{eq} = e\Phi|_{y=0}$, and $\xi_h|_{y=L_y} + \xi_{eq} = 0$. For $\Phi$, $(\partial_x \Phi)|_{x=0} = (\partial_x \Phi)|_{x=L_x} = 0$, $\Phi|_{y=L_y} = 0$, and $V - \Phi|_{y=0} - R_s L_{z,VO2} \int_0^{L_x} e(j_{hy} - j_{ey})|_{y=0} dx = 0$. The boundary conditions for $\Phi$ at $x = 0$ and $x = L_x$ reflect that the component of the electric field perpendicular to the boundaries is zero, otherwise the charges would accumulate following the electric field at the boundaries to screen the electric field to zero. The boundary condition for $\Phi$ at $y = L_y$ reflects that the boundary is connected to the ground. The complex boundary condition for $\Phi$ at $y = 0$ takes into account the voltage drop across the series resistor due to the current flowing through the resistor. In the finite element method, this non-standard boundary condition can be realized by the Lagrange multiplier method. For $T$, $-\kappa \nabla T \cdot \mathbf{n} = h(T - T_s)$ at all boundaries accounting for heat dissipation through the sides of the system, where $\mathbf{n}$ is the unit vector perpendicular to each boundary pointing outward.

The polycrystalline nature and local non-uniformities in oxygen stoichiometry or strain of the VO2 sample are accounted for by using a spatially-correlated random $T_c$-variation field with $\Delta T_c$ ranging from -20 K to 0 K with a correlation length of 25 nm. This approximate correlation length scale is based on atomic force microscopy (AFM) measurements of the film morphology (see Fig. S6). $T_c$ variation is realized by the variation of the "Curie-Weiss" temperatures of both $\mu$ and $\eta$ by the same amount. $T_s$ is chosen such that $T_c - T_s = 45$ K, which reflects the temperature difference in the experiment. The size of the system is chosen to be $L_x = 120$ nm, $L_y = 60$ nm, and $L_{z,VO2} = 60$ nm, so that $L_y/L_xL_{z,VO2}$, and therefore the resistance of the channel is similar to that in the experiment. The applied voltage $V$ is set to 0.0222 V providing an electric field similar to the experiment (=3.7 kV cm$^{-1}$). $R_s$ is set to 5 kΩ, to match the experimental value.

We use the finite element method to solve the partial differential equations for $\mu, \eta, n, p, \Phi$, and $T$. The simulated VO$_2$ resistance and phase fractions as functions of the time show an incubation time for the IMT of ~20 µs (Fig. S7), in reasonable agreement with the experiment. Furthermore, the simulation shows that the VO$_2$ temperature reaches close to $T_c$ at the onset of the IMT. These findings clarify the dominant role of the Joule heating in producing the finite incubation period observed in the experiments (Fig. 3D-F). Additional supporting calculations are presented below.

**Supplementary Notes**

Heat transfer coefficient

The phase-field simulations described above require an estimate of the effective heat transfer coefficient $h$. To estimate this, we employ a resistor network model for heat conduction into the Si substrate. Lateral heat transport in the VO$_2$ and SiN membranes is modeled using expressions for the radial thermal resistance,

$$R_{VO2} = \frac{\log(r_2/r_1)}{2\pi L_{z,VO2}\kappa_{VO2}}, \quad \text{(S11a)}$$

$$R_{SiN} = \frac{\log(r_2/r_1)}{2\pi L_{z,SiN}\kappa_{SiN}}, \quad \text{(S11b)}$$

where $\kappa_i$ is thermal conductivity, $L_{z,i}$ is thickness, and $r_2$ & $r_1$ are the external and internal radii. Note that the interfacial thermal resistance between VO$_2$ and SiN is negligible compared to the lateral thermal transport resistance through the membrane (by ~5 orders of magnitude) and is ignored here. We model the Pt metal leads as linear resistors with thermal resistance given by,

$$R_{Pt} = \frac{L_{Pt}}{L_xL_{z,Pt}\kappa_{Pt}}. \quad \text{(S12)}$$

To a first approximation, $r_2 \approx W_0/2, r_1 \approx \sqrt{L_xL_y}/2, L_{Pt} \approx (W_0 - L_y)/2$, where $W_0$ is the width of the suspended SiN membrane. The effective heat transfer coefficient $h$ is given by:

$$h = \frac{R_{VO2}^{-1} + R_{SiN}^{-1} + 2R_{Pt}^{-1}}{L_xL_y} \quad \text{(S13)}$$

Using $L_x = L_y = 20$ µm, $W_0 = 200$ µm, $L_{z,VO2} = 60$ nm, $L_{z,SiN} = L_{z,Pt} = 50$ nm, $\kappa_{VO2} = 4$ Wm$^{-1}$K$^{-1}$, $\kappa_{SiN} = 3$ Wm$^{-1}$K$^{-1}$, and $\kappa_{Pt} = 60$ Wm$^{-1}$K$^{-1}$, we estimate $h \approx 6000$ Wm$^{-2}$K$^{-1}$.

Analytical expression for the incubation time

As discussed above, the phase-field simulations suggest that the incubation phenomenon observed in experiments (Fig. 3D-F) can largely be explained as a Joule heating process. To further evaluate this, we derive a simple analytical expression. Assuming uniform heating and temperature rise in a 2D film, the heat diffusion equation can be written as:

$$[L_x L_y L_{z,VO2}] C_v \frac{\partial T}{\partial t} = \frac{[L_x L_{z,VO2}] \sigma V^2}{L_y} - [L_x L_y] h (T - T_s), \tag{S14}$$

Solving this equation gives,

$$T(t) = \frac{1}{\alpha} [\beta - (\beta - \alpha T_s) \exp(-\alpha t)], \tag{S15}$$

where $\alpha = \frac{h}{L_{z,VO2} C_v}$, and $\beta = \frac{\sigma V^2}{L_y^2 C_v} + \frac{h T_s}{L_{z,VO2} C_v}$. Setting $T = T_c$ at $t = t_{inc}$, we evaluate the incubation time:

$$t_{inc} = \frac{-L_{z,VO2} C_v}{h} \log \left[ 1 - \frac{h(T_c - T_s) L_y^2}{\sigma V^2 L_{z,VO2}} \right]. \tag{S16}$$

Using $h$ = 6000 Wm$^{-2}$K$^{-1}$, $T_c - T_s$ = 45 K, $\sigma \approx$ 170 Sm$^{-1}$, $C_v$ = 3.6 MJm$^{-3}$K$^{-1}$, we estimate $t_{inc}$~20 μs for $V$ = 5 V. This is comparable to experimental and phase-field simulation results and supports the suggestion that incubation can be explained as a thermal phenomenon.

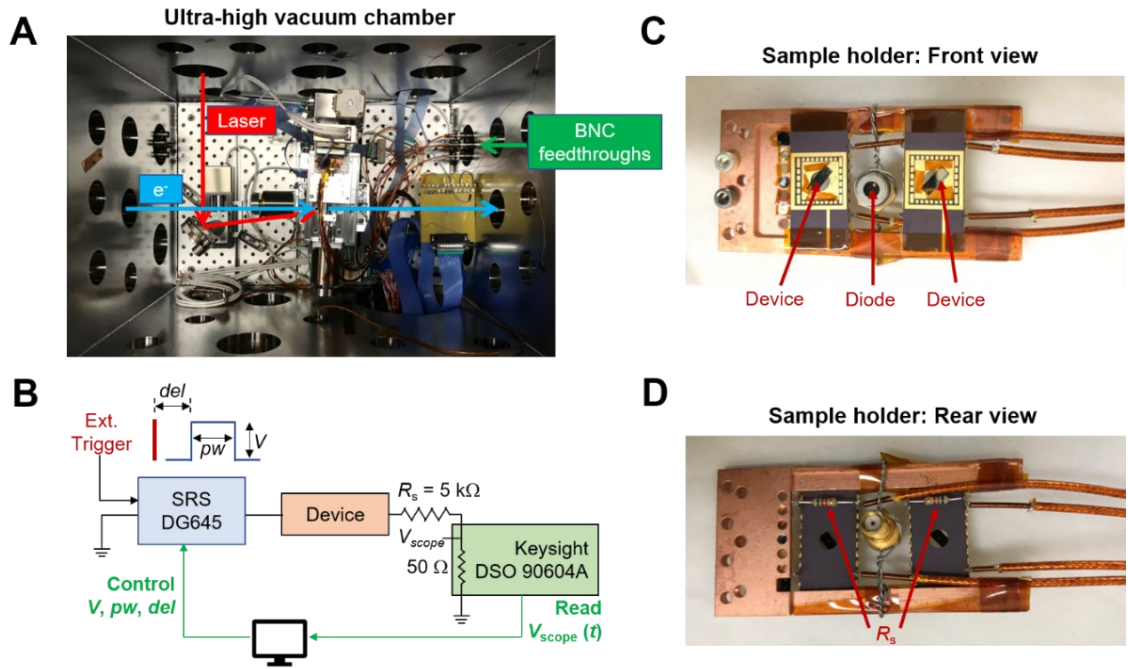

**Fig. S1.**

(A) Photograph of the inside of the ultrahigh vacuum (UHV) chamber showing laser and electron beam paths. (B) Schematic of electrical layout showing the various instruments and control parameters. (C) Front view and (D) rear view of the sample holder, showing two $VO_2$ devices mounted on chip carriers and connected to UHV compatible BNC cables. A fast diode enables precise timing between electrical pump, optical pump and MeV electron probe. Also seen in the rear view are 5 kΩ series resistors (labelled "$R_s$") that are soldered to the ground side of each $VO_2$ device.

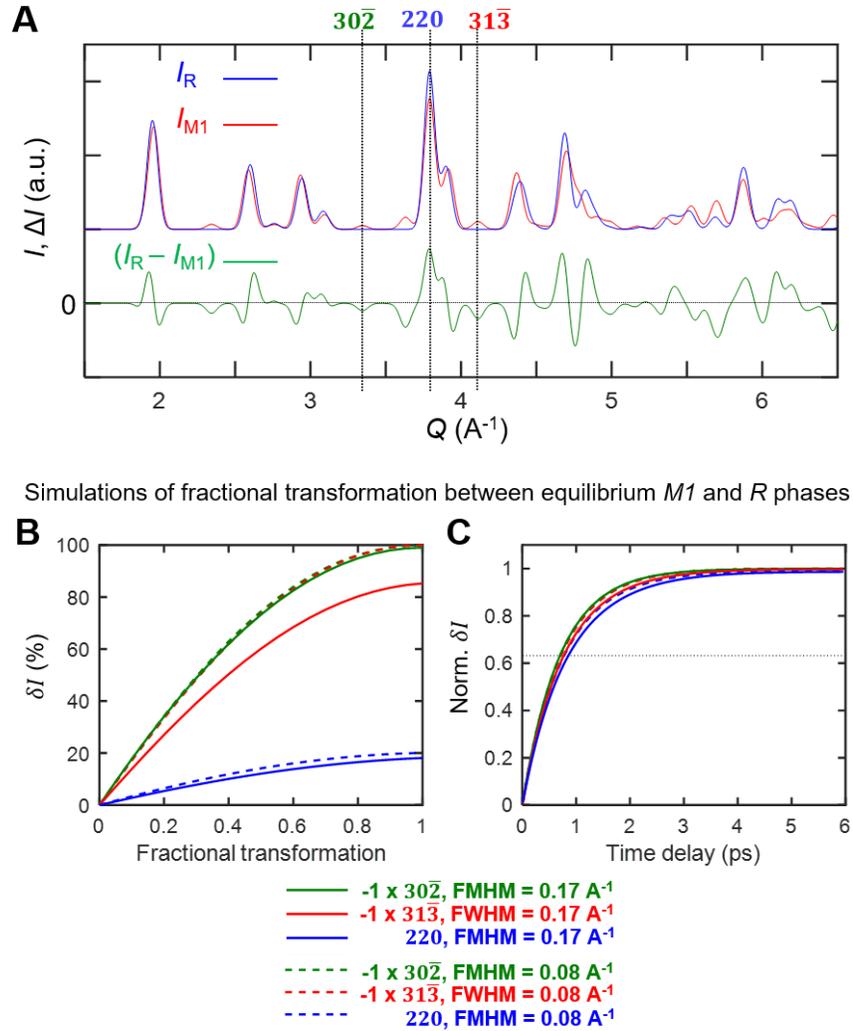

**Fig. S2.**
(A) Calculated diffraction patterns of the equilibrium monoclinic *M1* (red) and rutile *R* (blue) structures. Also shown is the intensity difference (green), $\Delta I = I_R - I_{M1}$. Three peaks of interest are labeled: $(30\bar{2})$, $(31\bar{3})$ and $(220)$. (B) Calculated differential intensity change ($\delta I = \Delta I / I$) versus fraction of *M1* transformed to *R*, assuming the phase transformation proceeds via linear movement of atoms between the initial (in *M1*) and final (in *R*) coordinates. (C) Simulated optical response based on the phase fraction dependent intensity change calculated in (B), and fixing the exponential time constants of $\delta I_{30\bar{2}}$ and $\delta I_{31\bar{3}}$ to 0.7 ps based on experimental measurements. The time constant of $\delta I_{220}$ is found to be < 0.9 ps. Clearly, this model, which only considers the two equilibrium phases, cannot explain the experimental observation that $\delta I_{220}$ changes on a ~4× slower timescale compared to $\delta I_{30\bar{2}}$ and $\delta I_{31\bar{3}}$ (see Fig. 3H). In (B) and (C), solid and dashed lines denote calculations based on a Gaussian broadening with FWHM of 0.17 Å$^{-1}$ and 0.085 Å$^{-1}$, respectively. The former is close to the *Q*-resolution of our UED system(*24, 37*).

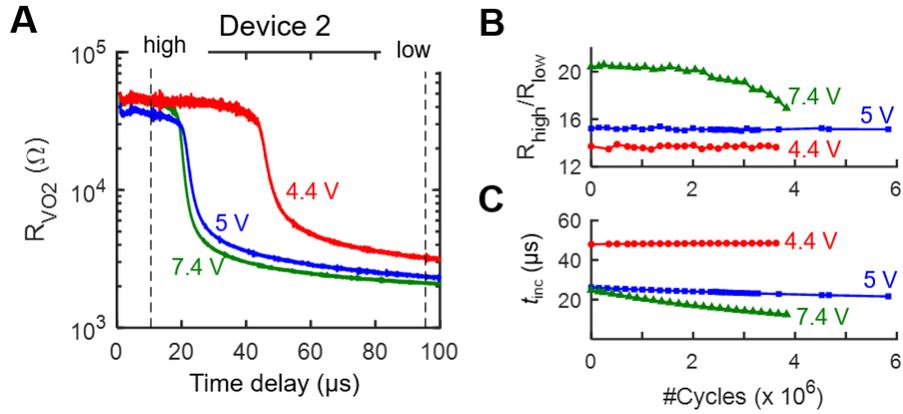

**Fig. S3.**
(A) Transient electrical resistance of Device 2 (channel width = length = 20 μm) versus delay time for three voltages 4.4, 5 and 7.4 V. (B) Resistance switching ratio and (C) incubation time (defined as the time delay at which $R_{VO2} = 10$ kΩ) versus cycle number. Note that although the data at 7.4 V show some degradation during cycling, the extracted trends for the UED structural and phase dynamics are robust, as shown in Fig. S5D-F.

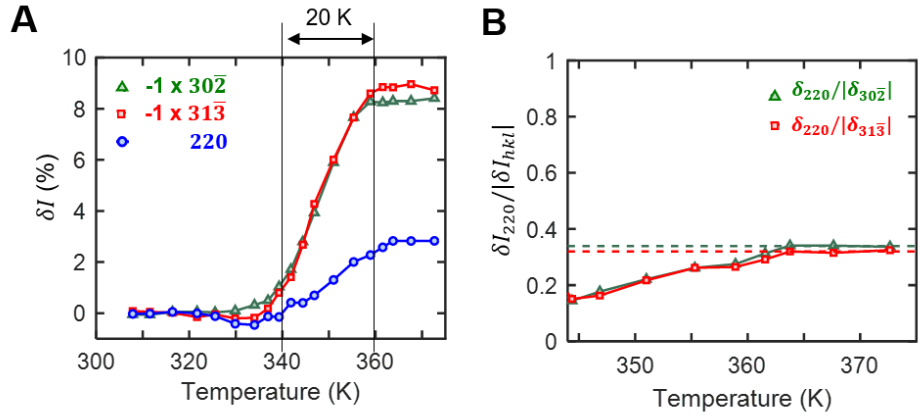

**Fig. S4.**

(A) Differential intensity change ($\delta I = \Delta I / I$) of the $(30\bar{2})$, $(31\bar{3})$ and $(220)$ peaks versus sample temperature under equilibrium heating conditions. The structural transition begins near 340 K and has a width ~20 K. This broadened structural transition is consistent with previous reports on polycrystalline films(*26*). (B) Ratio of differential intensity changes, $\delta I_{220}/|\delta I_{30\bar{2}}|$ and $\delta I_{220}/|\delta I_{31\bar{3}}|$ versus temperature. The maximum values of these ratios at 370 K are 0.32 and 0.34, indicated by the red and green dashed lines, respectively.

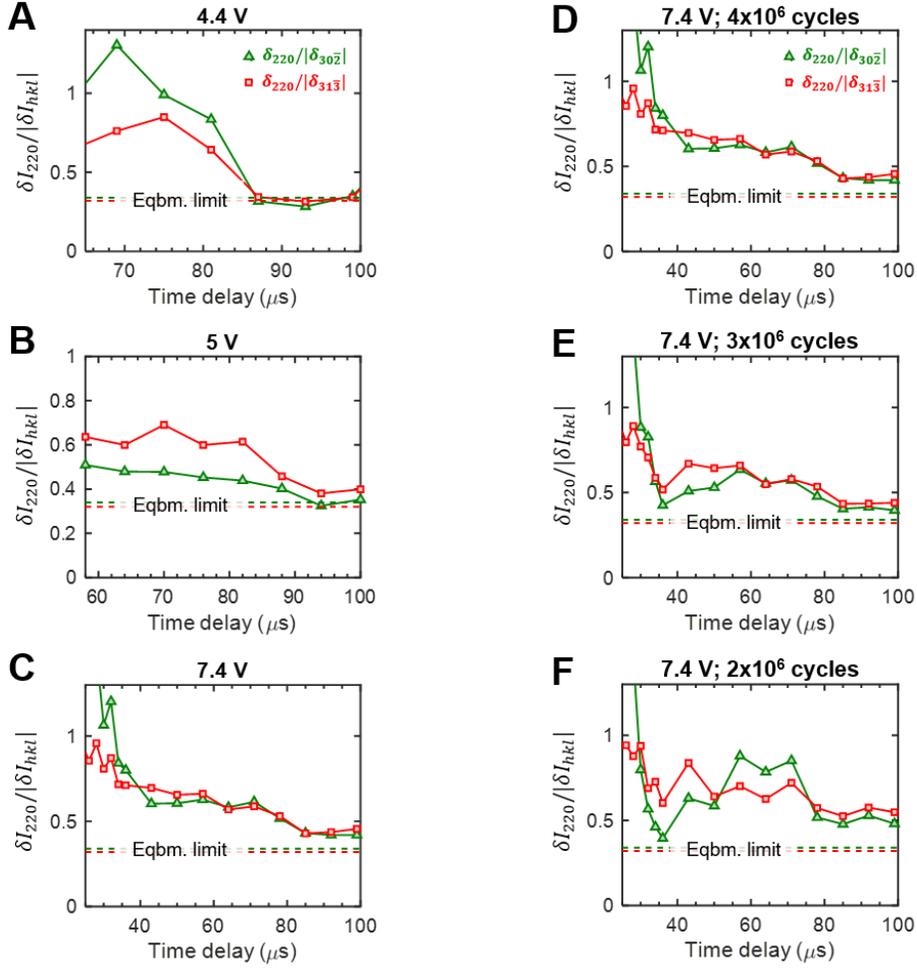

**Fig. S5.**
(A) – (C) Ratio of differential intensity changes $\delta I_{220}/|\delta I_{30\bar{2}}|$ and $\delta I_{220}/|\delta I_{31\bar{3}}|$ versus delay time for voltages 4.4, 5 and 7.4 V on Device 2 (see Fig. 3D-F). Dashed lines indicate the maximum values of these ratios corresponding to the equilibrium $M1 \rightarrow R$ transformation measured under quasi-static heating. In each case, the ratios are larger than can be explained by the equilibrium transformation and point towards the existence of a non-equilibrium metallic monoclinic (*mM*) phase. Furthermore, these ratios decrease with time and converge to the equilibrium values at ~100 μs showing that the *mM* phase exists transiently and eventually converts to the *R* phase. (D) – (F) Ratio of differential intensity changes $\delta I_{220}/|\delta I_{30\bar{2}}|$ and $\delta I_{220}/|\delta I_{31\bar{3}}|$ versus delay time for 7.4 V, averaging data taken over different number of cycles, $4\times10^6$ (entire data set), $3\times10^6$, and $2\times10^6$. Although the data get noisier with fewer cycles, the main trends remain - the ratios are larger than the equilibrium values and decrease at longer times towards the equilibrium values. In (A) – (F), we perform 3-point adjacent averaging to smoothen the data for better visualization.

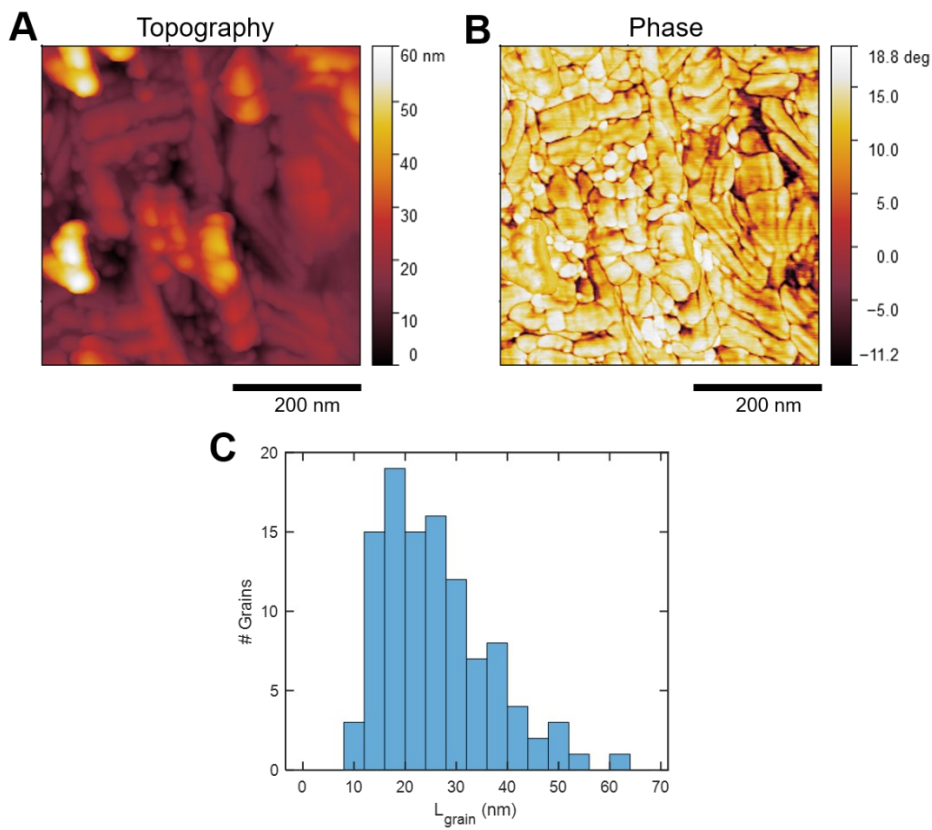

**Fig. S6.**

Atomic force microscope (AFM) image of VO$_2$ film (A) topography and (B) phase, clearly showing the polycrystalline grain structure. (C) Grain size distribution extracted from the AFM phase image.

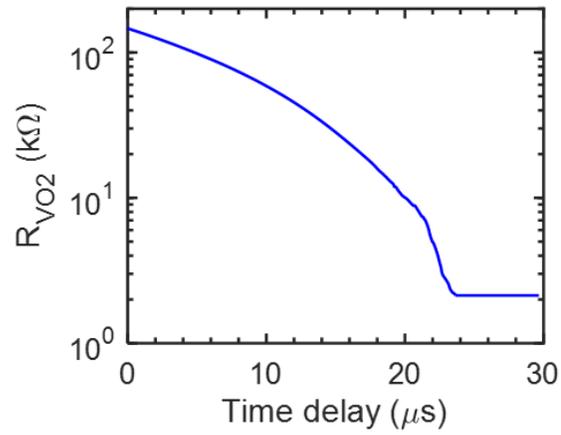

**Fig. S7.**
Phase-field simulation results showing the time-dependent electrical resistance corresponding to the results shown in Fig. 4C and Movie S1. An incubation time of ~20 μs is observed.

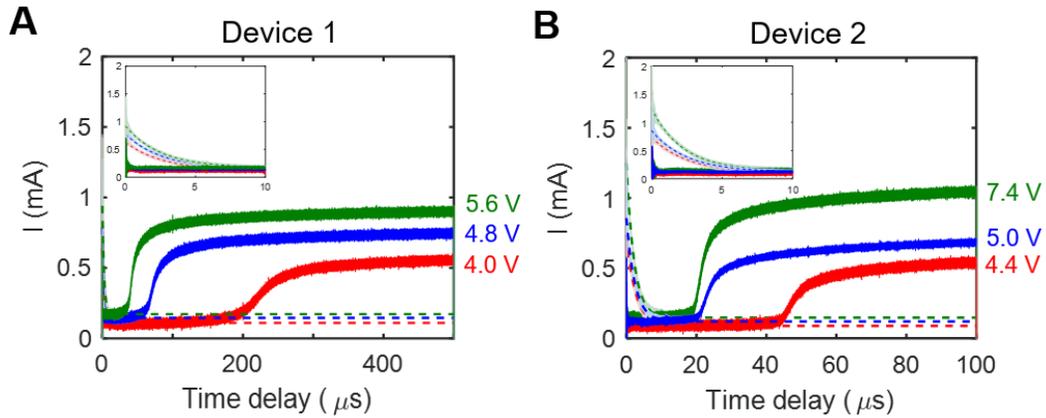

**Fig. S8.**
Raw current versus time traces showing the transient electrical characteristics of (A) Device 1 (channel width = 40 μm, length = 20 μm; structural data in Fig. 3A-C) and (B) Device 2 (channel width = length = 20 μm; structural data in Fig. 3D-G) for different applied voltages. The raw traces show a current spike at zero time delay, which decays exponentially within ~10 μs. This capacitive component is removed by fitting a single exponential function (see insets) and subtracting it from the total current to get the resistive component. This corrected current trace is used to extract the $VO_2$ channel resistance $R_{VO2}$ which is plotted in Fig. 3D-F and Fig. S3A.

**Movie S1.**

Phase-field simulations showing the phase evolution after the application of a step voltage. The simulation domain is 60 nm × 120 nm. Electric field (= 3.7 kV cm$^{-1}$) is applied along the vertical direction at *t* = 0, as shown in Fig. 4A.

Filename: 'Movie_S1.mp4'